# Ultrafast laser driven ferromagnetic-antiferromagnetic skyrmion switching in 2D topological magnet

Kaiying Dou, Wenhui Du, Zhonglin He, Ying Dai[*], Baibiao Huang, and Yandong Ma[*]

School of Physics, State Key Laboratory of Crystal Materials, Shandong University, Shandanan Str. 27, Jinan 250100, China

E-mail: daiy60@sdu.edu.cn (Y.D.); yandong.ma@sdu.edu.cn (Y.M.)

## Abstract

Light-spin coupling is an attractive phenomenon from the standpoints of fundamental physics and device applications, and has spurred rapid development recently. Whereas the current efforts are devoted to trivial magnetism, the interplay between light and nontrivial spin properties of topological magnetism is little known. Here, using first principles, rt-TDDFT and atomic spin simulations, we explore the evaluation of topological spin properties of monolayer $CrInSe_3$ under laser, establishing the ultrafast ferromagnetic-antiferromagnetic skyrmion reversal. The physics correlates to the laser-induced significant spin-selective charge transfer, demagnetization, and time-dependent magnetic interactions. Especially, an essential switching from ferromagnetic to antiferromagnetic exchange is generated under light irradiation. More importantly, dynamics of topological magnetic physics shows that this process accompanies with the evaluation of topological magnetism from ferromagnetic to antiferromagnetic skyrmions, manifesting intriguing interplay between light and topological spin properties. Our letter provides a novel approach toward the highly desired ultrafast control of topological magnetism.

# Introduction

Light-spin coupling is a booming topic in the field of condensed-matter physics and material science [1-4]. In this context, the ultrafast control of magnetic properties can be realized with laser pulse, which reaches a time scale of a few picoseconds or even femtoseconds [4-7]. This phenomenon holds great potential for advancing fundamental physics and promising a wide range of technological applications with high speed, including all-optical switching, high-speed data processing and opto-spintronics devices [8-10]. Especially with the discovery of long-range magnetism in monolayer $CrI_3$, much effort has been dedicated to exploring the interplay between ultrafast laser and various magnetic properties in two-dimensional (2D) materials [11,12]. For example, in recent studies the magnetic phase transitions under laser pulse have been observed in 2D systems of $CrCl_3/CrBr_3$ [10], $RuCl_3$ [13], $MnPS_3/MnSe_2$ [14] and $Cr_2CCl_2/MnS_2$ [15]. In addition, another study shows that the reversal of magnetic anisotropy can be observed experimentally via imposing ultrashort laser pulse in 2D antiferromagnet $NiPS_3$ [16]. All these works prove 2D magnetic materials a compelling category of light-spin coupling medium.

Different from trivial magnetism, spins in topological magnetism (e.g., skyrmion, bimeron) can form topologically protected chiral magnetic quasi-particles with whirling spin textures [17-19]. Stable topological magnetism emerges in various magnetic materials due to the competition of Heisenberg exchange and Dzyaloshinskii-Moriya interaction (DMI) [20]. Physically, akin to trivial magnetism, ultrafast laser can also couple with nontrivial spin properties of topological magnetism, delivering the possibility of ultrafast control of topological magnetic physics. Once achieved, it makes a significant step toward high-performance spintronic devices based on 2D topological magnetism. Through highly desirable, the interplay between light and nontrivial spin properties of topological magnetism remains largely unexplored and less known [21,22]. Especially, the ultrafast laser control of ferromagnetic-antiferromagnetic (FM-AFM) topological spin switching, one of the most long-sought phenomena in topological magnetism, has never been reported.

In this letter, by combining first principles, real-time time-dependent density functional theory (rt-TDDFT) and atomic spin model simulations, we investigate the interplay between ultrafast laser pulse and topological spin properties in monolayer $CrInSe_3$, proving the laser-induced reversal of FM-AFM skyrmions. Our results show that such coupling is attributed to the laser-induced significant spin-selective charge transfer, demagnetization and time-dependent magnetic interactions. In particularly, the magnetic exchange experiences an essential switching from FM to AFM under light irradiation. More remarkably, the time resolved spin textures of monolayer $CrInSe_3$ further suggests that this process is accompanied with reversing topological magnetism from FM to AFM skyrmions. This indicates the efficient interplay between light and topological spin properties in monolayer $CrInSe_3$. Our letter highlights the importance of light in the realization of ultrafast control of topological magnetism.

## Results and Discussion

**Fig. 1** presents the crystal structure of monolayer CrInSe$_3$ [23], which exhibits the space group *P*3m1 and consists of five atomic layers. Each unit cell contains one Cr, one In and three Se atoms, and these atoms are stacked in the sequence of Se-In-Se-Cr-Se. The Cr atom coordinates with six Se atoms, forming the distorted octahedral geometry. As a result, the inversion symmetry is broken. The lattice constants of monolayer CrInSe$_3$ are optimized to be *a* = *b* = 3.92 Å, consistent with the previous work [23]. To assess the stability of monolayer CrInSe$_3$, the phonon spectra is calculated. As shown in **Fig. S1a**, only a tiny imaginary frequency (< 10/cm$^{-1}$) is observed around the Γ point, which can be attributed to the long-range periodic rippling on the top of local corrugations [24,25]. This indicates that the monolayer CrInSe$_3$ is dynamically stable. Its thermal stability is further estimated by performing ab initio molecular dynamics (AIMD) simulations. The slight free-energy fluctuation and well-defined structure shown in **Fig. S1b** suggest the thermal stability of monolayer CrInSe$_3$.

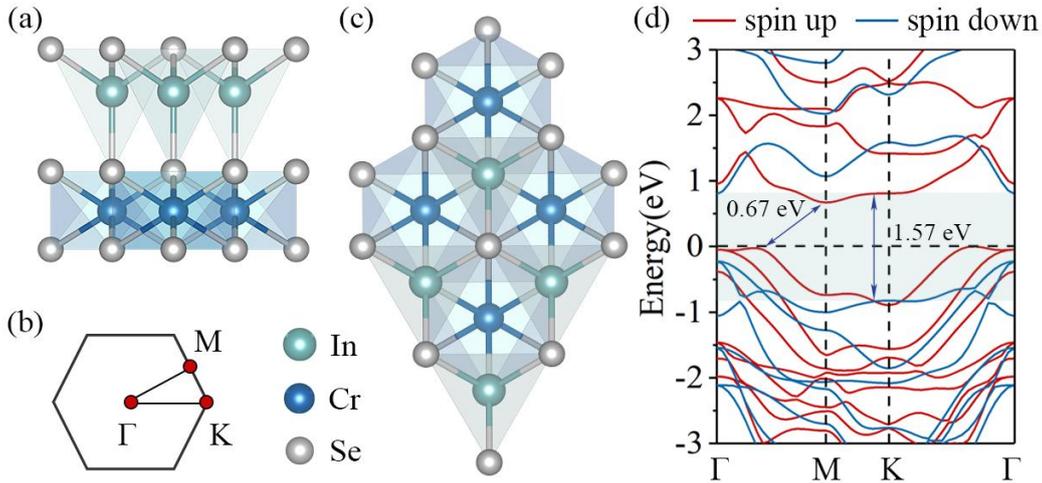

**Fig. 1** (a) Side and (c) top views of the crystal structure of monolayer CrInSe$_3$. (b) Schematic diagram of 2D Brillouin zone. (c) Spin-polarized band structures of monolayer CrInSe$_3$. The Fermi level is set to the valance band maximum. The indirect band gap and maximum direct band gap are illustrated in (c).

The valence electronic configuration of Cr atom is $3d^54s^1$. In monolayer CrInSe$_3$, each Cr atom donates three electrons to the coordinated Se atoms, yielding the valence electronic configuration of $3d^34s^0$. Under distorted octahedral coordination environment, Cr-3*d* orbitals roughly split into two groups: high-lying $e_g(d_{xy}, d_{x^2-y^2})$ and low-lying $t_{2g}(d_{xz}, d_{yz}, d_{z^2})$. The three electrons of Cr atom half occupy $t_{2g}$ orbitals, which would result in a magnetic moment of 3 μ$_B$. Our calculations confirm this analysis and show that the magnetic moment is mainly distributed on Cr atom. **Fig. 1d** shows the spin-polarized band structure of monolayer CrInSe$_3$, which exhibits an indirect band gap of 0.67 eV. Therefore, monolayer CrInSe$_3$ is a 2D magnetic semiconductor.

To further uncover the magnetic properties of monolayer CrInSe$_3$, we construct a 2D classical

Heisenberg model [26,27], wherein the spin Hamiltonian can be expressed as

$$H = -J \sum_{<i,j>} (\mathbf{S}_i \cdot \mathbf{S}_j) - \sum_{<i,j>} \mathbf{d}_{xy} \cdot (\mathbf{S}_i \times \mathbf{S}_j) - K \sum_i (S_i^z)^2 - mB \sum_i S_i^z.$$

Here, $\mathbf{S}_{i/j}$ is the spin magnetic moment on each atomic site. $J$, $\mathbf{d}_{xy}$ and $K$ characterizes the strengths of Heisenberg exchange coupling, DMI and single ion anisotropy (SIA), respectively. The last term is the Zeeman energy. The Heisenberg exchange interaction $J$ is calculated to be 26.55 meV. It suggests the FM exchange interaction is favorable between the nearest neighboring magnetic moments. Based on the Moriya's rule [28], DMI for the nearest-neighboring Mn atoms is perpendicular to their bond. As the out-of-plane component has no contributions on the spiral arrangement of magnetic atoms [29,30], we only consider its in-plane component $\mathbf{d}_{xy}$. The $|\mathbf{d}_{xy}|$ is calculated to be 2.03 meV. Furthermore, the SIA parameter $K$ is calculated to be 0.24 meV, which implies the preference of out-of-plane magnetization.

Based on the first-principles parameterized Hamiltonian, we conduct atomic spin model simulations to explore the spin textures of monolayer $CrInSe_3$. The spin texture is displayed in **Fig. 2a**, wherein the skyrmions with labyrinth domains is observed, consistent with the previous work [23]. In fact, this is consistent with its proper ratio of $|\mathbf{d}_{xy}/J| = 0.076$, which is near the typical range of 0.1-0.2 for hosting magnetic skyrmions [23,29]. When applying an external magnetic field of 0.2 T, the labyrinth domains are fragmented and the size of skyrmions shrinks. With further enhancing the magnetic field to 0.4 T, the size of the skyrmions continues to decrease while the density increases. In view the coexistence of semiconducting character and topological magnetism, monolayer $CrInSe_3$ is an promising system to investigate the interplay between light and the nontrivial topological magnetism.

We then perform rt-TDDFT calculations to explore the spin dynamics of topological magnetism under ultrafast laser pulses in monolayer $CrInSe_3$. For initiating the spin dynamics, we impose a linearly polarized (in-plane polarization) pulse along $x$ axis to monolayer $CrInSe_3$. From **Fig. 1d**, we can see that the maximum direct band gap of monolayer $CrInSe_3$ is 1.57 eV. By considering this fact, the laser pulse harbors a photon energy of 1.63 eV, a full width at half-maximum of 6.04 fs, and a fluence rate of 13.95 mJ/cm$^2$. The time-dependent dynamics of magnetic moments on Cr and In atoms are shown in **Fig. 2b**. Obviously, significant time-dependent changes of magnetic moments are observed for Cr and In atoms. Specifically, the demagnetization occurs for Cr atom, which begins at ~7 fs and then the magnetic moment stabilizes at ~15 fs. After ~15 fs, the inertia does not further promote the demagnetization, and the magnetic moment reaches at dynamic equilibrium with slight oscillation until the end of our simulation at ~36 fs. This results in the change of delocalized moment of $\Delta m = $ ~ 0.15 $\mu_B$ for Cr atom. The time scale of the whole demagnetization corresponds to the duration of the imposed laser pulse, which is similar to the cases of $MnSe_2$ and $RuCl_3$ [14,31]. While for In atom, it starts to magnetize from ~7 fs and then the magnetic moment stabilizes at ~15 fs. In detail, In atom gains a magnetic moment of ~ 0.06 $\mu_B$. Different from the case of Cr and In, as shown in **Fig. 2c**, the magnetic moments of Se2 and Se3 atoms, which originally exhibit negligible values, remain almost constant with minimal fluctuations under ultrafast laser. While for Se1 atom, it has a small original magnetic moment, which increases slightly under the effect of laser pulse. Obviously, the laser-induced demagnetization of Cr atom, magnetization of In and negligible moment changes for Se imply the significant optical intersite spin-transfer between Cr and In atoms.

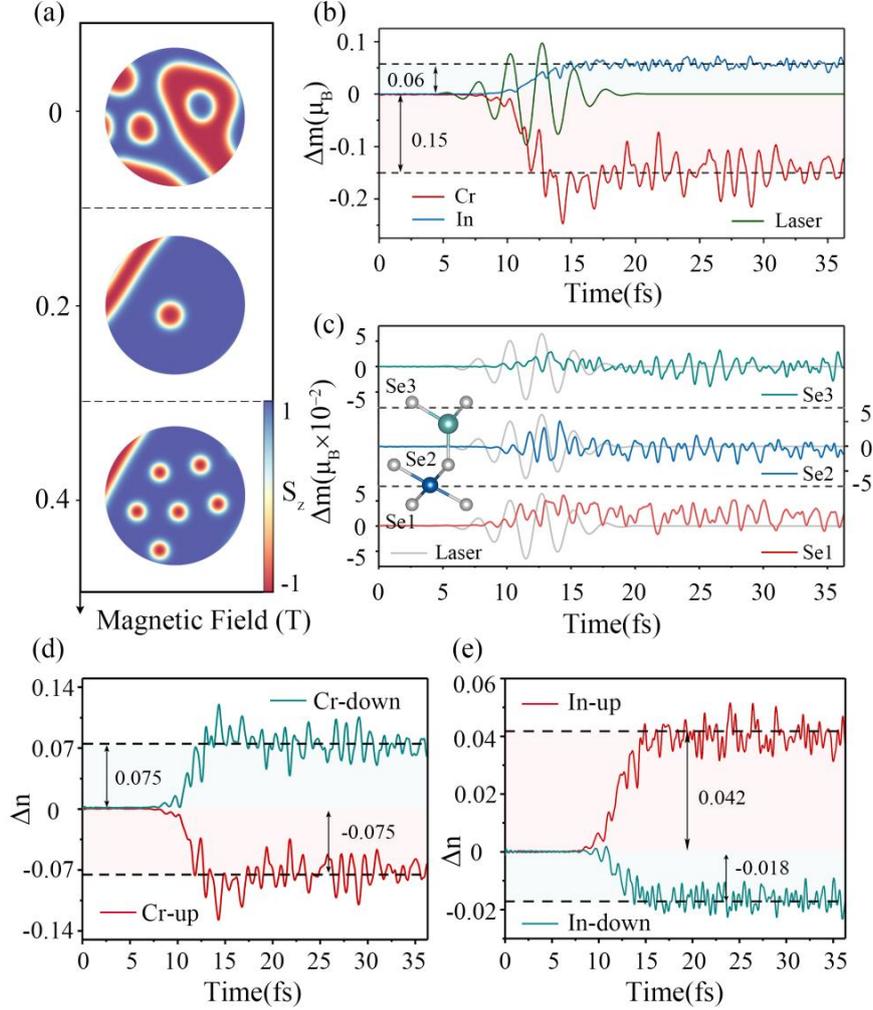

**Fig. 2** (a) Spin textures of monolayer CrInSe$_3$ with and without external magnetic fields. (b) Time evolution of the variation on localized magnetic moments of (b) Cr/In and (c) three Se atoms in monolayer CrInSe$_3$ under ultrafast laser pulse, including the diagram of applied laser pulse. Time evolution of the variation of occupation electron on the two spin channels as a function of time for (d) Cr and (e) In atoms.

Such demagnetization and magnetization features driven by ultrafast laser in monolayer CrInSe$_3$ correlates to the time-dependent electron occupation [$\Delta n(t)$] of the majority and minority channels in Cr and In atoms. As shown in **Fig. 2d** and **e**, the majority (spin-up)/minority (spin-down) channel of Cr atom loses/gains electrons at a relatively rapid rate in the time window of 7 ~ 15 fs, with the occupation loss/gain being approximately 0.075. It leads to the demagnetization of Cr atom since the magnetic moment is directly proportional to the electron occupation of the majority and minority channels. For In atom, its majority channel sees an occupation gain of 0.042, and occupation loss of 0.018 is observed for its minority channel within the time window of 7 ~ 15 fs.

This results in a photo-generated magnetic moment 0.06 $\mu_B$ for initially non-magnetized In atom. For Se atoms, as shown in **Fig. S2**, both spin channels almost neither gain or lose under laser irradiation. It confirms that, with applying laser pulse on monolayer CrInSe$_3$, spin-up electrons from Cr atom are partially transferred to In atoms, while spin-down electrons from In atom are correspondingly transferred to Cr atom.

In addition to magnetic moments, the magnetic interactions among magnetic moments in monolayer CrInSe$_3$ would also be influenced by the ultrafast laser pulse. In this sense, we next investigate the time-dependent magnetic interactions under ultrafast laser pulse. To set the stage, we first study the time-dependent total density of states (DOS) of monolayer CrInSe$_3$ under ultrafast laser pulse to estimate the excitation of electrons in momentum space. Taking the excitation cases of 2.42 and 24.19 fs as examples, the DOS of monolayer CrInSe$_3$ under laser are presented in **Fig. 3a** and **b**. It can be seen from **Fig. 3a** that at 2.42 fs, there is no electron in the conduction band, and the valence band is fully occupied, indicating that the electron-hole (e-h) pair is not generated yet. At 24.19 fs, electrons are excited to the conduction band, which suggests that laser excitation generates e-h pairs in monolayer CrInSe$_3$. Evidently, the region of DOS over the conduction band (dark regions in **Fig. 3b**) represents the density of excited electrons. On the other hand, as it is an isolated system, the excited electrons in conduction band also corresponds to the generated holes in valence band [32]. In view of these facts, the density of laser-generated e-h pairs at each time point can be obtained from the corresponding DOS of monolayer CrInSe$_3$.

Following the above analysis, the time-dependent variation of the density of e-h pairs is estimated and presented by the green line and red star symbols in **Fig. 3c**. This, along with the time-dependent demagnetization of Cr atom (blue line in **Fig. 3c**), can roughly divide the laser-excited process into three stages. In detail, before ~9.68 fs (stage I), the magnetic moment remains unchanged and electronic excitation is not occurred with no generation of e-h pairs. Starting from 9.68 fs, the concentration of e-h pairs rises rapidly and then stabilize at approximately 16.94 fs (stage II). During this state, the demagnetization of Cr atom is significant. After 16.94 fs (stage III), the density of e-h pairs is stable and the demagnetization of Cr atom is finished, indicating the excitation of electrons by the ultrafast laser is completed. In the following, we take points ① of 2.42 fs, ②/③ of 12.10/14.51 fs and ④ of 24.19 fs as examples to represent these three stages to investigate the time-dependent magnetic interactions in monolayer CrInSe$_3$ under ultrafast laser pulse. The points ① and ④ correspond to the relatively stable states of the system before and after the laser pulse, respectively, while points ② and ③ represent the unstable and transient states during the process of laser pulse excitation.

Based on the time-dependent e-h pair density and PDOS, we employ constrained DFT and 2D classical Heisenberg model to explore the dynamics of magnetic interactions under ultrafast laser pulses in monolayer CrInSe$_3$. The time-dependent variations of magnetic parameters are displayed in **Fig. 3d-g**. For DMI interaction, as shown in **Fig. 3d**, $d_{xy}$ increases significantly from 2.03 meV at point ① to 3.48 meV at point ② and then to 6.47 meV at point ③. This corresponds to the increase of e-h pair density and demagnetization of Cr atom. By moving from point ③ to ④, as the e-h pair density and demagnetization of Cr atom reaches saturation, $d_{xy}$ undergoes a slight reduction and is stablished at point ④ of 4.62 meV. For the Heisenberg exchange, the variation trend is different from that of DMI. As shown in **Fig. 3e**, the exchange coupling parameter $J$ continuously decreases from 26.55 meV at point ① to 21.07 meV at point ②, to -5.19 meV at point ③ and finally to -22.00 meV at point ④. Particularly, the sign of $J$ changes from positive to negative between points ② and ③, *i.e.*, the FM-AFM switching is induced by laser pulse in monolayer CrInSe$_3$.

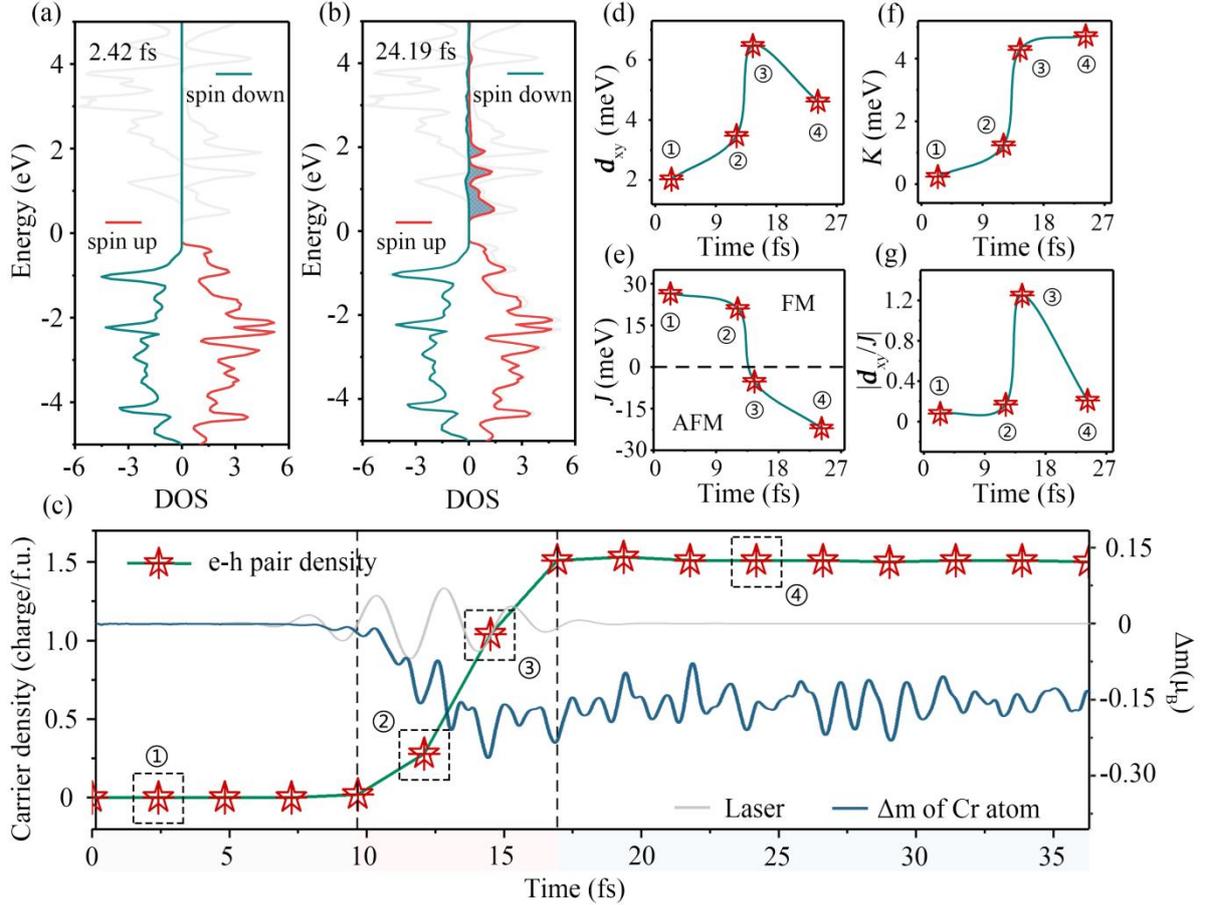

**Fig. 3** (a, b) Time-dependent DOS of monolayer CrInSe$_3$ at 2.42 and 24.19 fs, respectively. The intrinsic DOS is also plotted in (a, b) for comparison (grey lines). Red and green lines in (a, b) indicate the time-dependent DOS for spin-up and spin-down states, respectively. (c) Time evolution of e-h pair density in monolayer CrInSe$_3$ under ultrafast laser pulse. Blue and grey lines in (c) indicate the corresponding magnetic moment variation on Cr atom and the applied laser pulse, respectively. Dynamics of (d) DMI, (e) Heisenberg exchange interaction, (f) SIA and (g) $|d_{xy}/J|$ in monolayer CrInSe$_3$ under ultrafast laser pulse.

Such laser-induced FM-AFM switching in monolayer CrInSe$_3$ is attributed to the synthetic effect of direct orbital overlap and 90° superexchange coupling via Se1 bridge. For the intrinsic case, direct interaction is weak and superexchange coupling dominates the interaction due to the approximately 90° Cr-Se-Cr bonding angle and relatively localized magnetic moment of Se1 atom. Therefore, the FM coupling is favorable [33-35]. Triggered by the ultrafast laser, the magnetic moment on Se1 atom relatively is enhanced and the direct orbital overlap enlarges. This transfers the 90° superexchange coupling via Se1 bridge to favor AFM coupling. Consequently, as bridging atom mediating the coupling between two neighboring Cr atoms, the relative increase of magnetic moment on Se1 atom driven by the laser pulse plays a crucial role in the FM-AFM phase transition.

Similar to Heisenberg exchange coupling, the SIA also exhibits a monotonical variation, but with opposite trend. As shown in **Fig. 3f**, the SIA parameter $K$ increases from 0.24 meV at point ① to 1.24 meV at point ②, to 4.28 meV at point ③ and finally to 4.72 meV at point ④. In other words, the introduction of laser pulse significantly enhances the SIA strength of monolayer $CrInSe_3$. Based on the magnetic parameters obtained above, the absolute value of the ratio between DMI and Heisenberg exchange interaction is estimated to be $|d_{xy}/J|$ =0.08, 0.17, 1.25 and 0.21, respectively, at four points. It is worth to point that $0.1 < |d_{xy}/J| < 0.2$ is usually considered as a criterion to stabilize topological magnetism [23,29]. Since the $|d_{xy}/J|$ ratios at points ①, ② and ④ are close to this typical range, the corresponding states hold high possibility to host nontrivial topological magnetic physics. Recalling the states at points ①/② and ④ favoring FM and AFM coupling respectively, the ferromagnetic-antiferromagnetic (FM-AFM) topological spin switching, one of the most long-sought phenomena in topological magnetism, might be induced by the ultrafast laser in monolayer $CrInSe_3$. For the state at point ③, because its $|d_{xy}/J|$ ratio remarkably deviates from the criterion, topological magnetism would be disappeared.

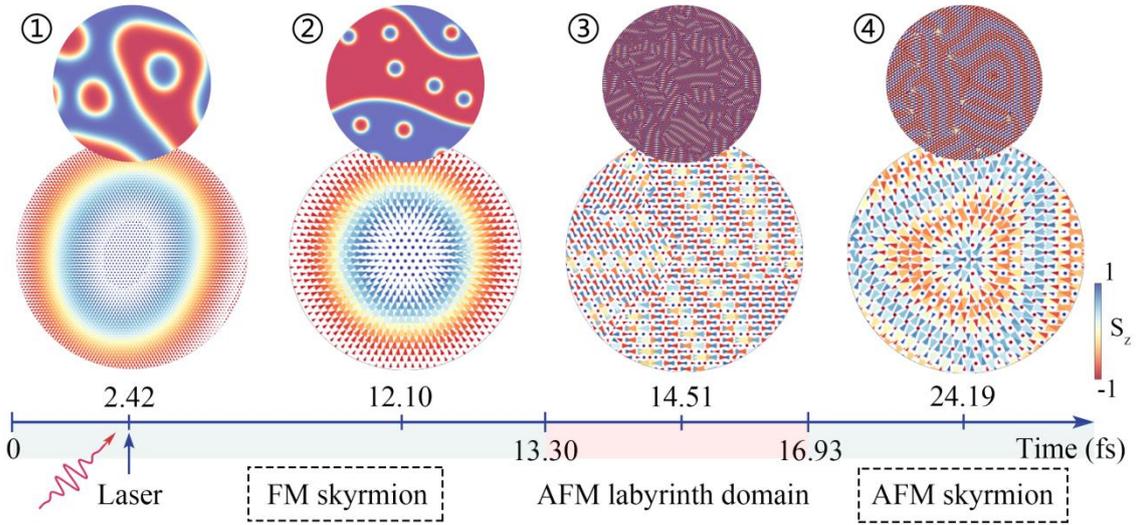

**Fig. 4** Dynamics of topological magnetic physics under ultrafast laser pulse in monolayer $CrInSe_3$.

To confirm these hypothesizes, atomic spin model simulations are performed to explore the dynamics of topological magnetic physics under ultrafast laser pulses in monolayer $CrInSe_3$. The corresponding results are presented in **Fig. 4**. At point ①, *i.e.*, without laser excitation, the spin texture exhibits a mixed configuration of FM skyrmions and labyrinth domains. The enlarged view of FM skyrmion region shows that the diameter of FM skyrmion is approximately 25 nm. Due to the influence of the irregular labyrinth domain, the FM skyrmions deviate from a perfectly circular shape, adopting an elliptical form instead. By moving from point ① to ②, as shown in the second panel of **Fig. 4**, FM skyrmions and labyrinth domains still coexist. However, as compared with point ①, driven by the laser-induced increase of DMI, the size of FM skyrmions at point ② decreases (with a diameter of approximately 11 nm) and the density increases. Meanwhile, the area occupied by FM skyrmions at point ② expands, while the labyrinth domains diminish. As a results, at point ②, the FM skyrmions come to dominate the majority of the area, which leads to that the FM skyrmions exhibit a standard circular shape.

With transforming to point ③, the FM-AFM transition occurs in monolayer $CrInSe_3$. And as displayed in the third panel of **Fig. 4**, only irregular AFM striped spin textures are observed, while the skyrmions disappear. In fact, this feature is expected from the excessively weakened exchange interaction and significantly enhanced DMI induced by laser excitation. With further moving to point ④, the reduction of DMI and enhancement of exchange interaction results in a transition from striped texture into a mixed state comprising labyrinth domains and skyrmions. Because the AFM exchange interaction is preserved, the resultant topological spin textures are AFM skyrmions. The localized spin texture containing the AFM skyrmion region is presented in an enlarged view in the last panel of **Fig. 4**. It can be seen that the AFM skyrmions exhibit a triangle-like shape. Moreover, attributing to the large DMI, the diameter of AFM skyrmion is approximately only 3.4 nm, which is significantly smaller than those reported in previously work [36,37]. The emergence of AFM skyrmion suggests that the FM-AFM skyrmion switching can be realized in monolayer $CrInSe_3$ by ultrafast laser pulse.

## Conclusions

To summarize, using first principles, rt-TDDFT and atomic spin model simulations, we systematically explore the coupling between ultrafast laser pulse and skyrmion physics in monolayer $CrInSe_3$. We reveal that significant spin-selective charge transfer, demagnetization and time-dependent magnetic interactions are induced by laser pulse. Furthermore, the transition from FM to AFM is observed under light irradiation. Most importantly, by uncovering the dynamics of topological magnetic physics under ultrafast laser pulse in monolayer $CrInSe_3$, we find that the reversal of topological magnetism from FM to AFM skyrmion emerges, guaranteeing the efficient interplay between light and topological spin properties. These phenomena greatly enrich the research on light-spin coupling and topological magnetism.

## Supplemental Material

Supplemental Material is available free of charge at ******.

(Note 1) Discussion on the stability of monolayer $CrInSe_3$; (Note 2) Time evolution of carriers for Se atoms; (Note 3) Supplemental illustration on time-dependent DOS; (Note 4) Supplemental analysis on Heisenberg exchange interaction; (Note 5) Magnetic parameters calculation details; (Note 6) Methods.

## Conflict of Interest

The authors declare no conflict of interest.

## Acknowledgments

This work is supported by the National Natural Science Foundation of China (Nos. 12274261 and 12074217), Taishan Young Scholar Program of Shandong Province, Young Students Fundamental Research Program (Qingmiao Program) of Shandong University (No. SDU-QM-Y2024002).